\documentclass[aps,pre,epsf,superscriptaddress,amsmath,amssymb,amsfonts,twocolumn,showpacs]{revtex4-1}
\usepackage{amsmath,amssymb,amsfonts,bm,comment}
\usepackage{graphics,float,tikz}
\usepackage{appendix}
\usetikzlibrary{arrows.meta,arrows}
\usepackage[caption=false]{subfig}
\usepackage[colorlinks=true,
	linkcolor=blue,
	filecolor=magenta,      
	urlcolor=blue,
	citecolor=blue]{hyperref}
\usepackage{physics,xparse,mathtools,isotope}
\usepackage[table-parse-only]{siunitx}
\usepackage{multirow,array,relsize,lipsum,longtable,tabularx}
\usepackage{dsfont}

\usepackage{amsthm}


\begin{document}
	\title{Using Entangled Generalized Coherent States for Photonic Quantum Metrology}
    \author{Madhura Ghosh Dastidar}
	\affiliation{Department of Physics, Indian Institute of Technology Madras, Chennai 600036, Tamil Nadu, India}
	
	\author{Aprameyan Desikan}
	\affiliation{Department of Physical Sciences, Indian Institute of Science Education \& Research Mohali Sector 81 SAS Nagar, Manauli PO 140306 Punjab, India}
	
	\author{Vidya Praveen Bhallamudi}
	\affiliation{Department of Physics, Indian Institute of Technology Madras, Chennai 600036, Tamil Nadu, India}
	
	\date{\today}
	
	\begin{abstract}
		

    Quantum metrology aims at achieving enhanced performance in measuring unknown parameters by utilizing quantum resources. Thus, quantum metrology is an important application of quantum technologies. Photonic systems can implement these metrological tasks with simpler experimental techniques. We present a scheme for improved parameter estimation by introducing entangled \textit{generalized} coherent states (EGCS) for photonic quantum metrology. These states show enhanced sensitivity beyond the classical and Heisenberg limits and prove to be advantageous as compared to the entangled coherent and NOON states. Further, we also propose a scheme for experimentally generating certain entangled generalized coherent states with current technology.

	\end{abstract}
	\maketitle
	
	\section{Introduction}\label{Introduction}
	With further advancements in quantum technologies, it has become necessary to progress in the precise measurement of parameters in quantum experimentation. Quantum metrology uses resources such as quantum entanglement and squeezing, that enhance the precision limits for measurement beyond the capabilities of classical frameworks~\cite{giovannetti2011advances,toth2014quantum}. With $N$ probes for the measurement of an unknown parameter $\phi$ using the classical framework, the minimum uncertainty in $\phi$ is limited by $\Delta\phi = 1/\sqrt{N}$ (shot-noise limit). However, a lower bound on the uncertainty can be achieved by entangled probes such that $\Delta\phi = 1/N$ (Heisenberg limit). 
 Now, the unknown parameter can be represented by a phase change in an interferometric setup and the probes are mapped to the mean photon number of the input state to the interferometer. 
	
	Photonic quantum metrology~\cite{dowling2014quantum,polino2020photonic} uses quantum interferometry which can enable sub shot-noise limit phase sensitivity, having applications in various fields such as quantum imaging and quantum frequency standards~\cite{giovannetti2004quantum}. Current research in this field aims at searching for optimal quantum states~\cite{hyllus2010not} at the input of the interferometer, phase measurement schemes~\cite{higgins2007entanglement} and considering noisy experimental conditions~\cite{dorner2009optimal,rosenkranz2009parameter}. Various reports in literature have discussed the advantage of certain entangled states of light such as  the NOON state~\cite{cooper2010entanglement} and entangled coherent states (ECS)~\cite{joo2011quantum} in phase estimation. While achieving large particle numbers for quantum-enhanced sensing is possible in atomic ensembles~\cite{esteve2008squeezing,riedel2010atom}, for photonic states this is quite challenging. For example, while NOON~\cite{bollinger1996optimal} and twin-Fock~\cite{holland1993interferometric} states have been proved to show enhanced sensitivity, their experimental generation remains limited to few photon states~\cite{nagata2007beating} as the efficiency decreases exponentially with the mean photon number~\cite{wang2016experimental}. Other quantum states such as entangled coherent~\cite{joo2011quantum} and superradiant photonic states~\cite{paulisch2019quantum}, possessing a continuous distribution for the mean photon number, require complicated experimental setups for generation. Thus, the quest for scalable entangled photonic states advantageous for quantum metrology remains open. 
    
     The advantage of the generalization of entangled coherent states for multi-parameter estimation~\cite{liu2016quantum} for quantum metrology has been explored in literature. However, the generalization of coherent states can also be performed by applying a displacement operator to non-zero photon number states. This type of generalization has been explored in theory~\cite{philbin2014generalized} and experiment~\cite{ziesel2013experimental}. 
     
     In this paper, we theoretically analyze using an entangled form of these generalized coherent states (GCS), namely, entangled coherent states (EGCS) for photonic quantum metrology. It is a two-mode entangled state of light where one more is vacuum and other mode is occupied by a GCS. We observe that, while EGCS can achieve the Heisenberg limit of phase sensitivity, under certain conditions these states can go even beyond this limit. Further, experimental generation of certain well-known states that are considered powerful resources for quantum metrology is quite tedious and yet, are limited by low fidelities and efficiencies. Thus, we also describe a simpler experimental scheme for generation EGCS.
     
     We start by introducing a general framework for calculating the minimum phase uncertainty for pure states in a photonic quantum metrology scheme and analytically derive that for EGCS in Sec.~\ref{SecII}. Next in Sec.~\ref{SecIII}, we compare the phase sensitivities (derived both numerically and analytically) obtained for EGCS with two common advantageous entangled states that show enhanced sensing $-$ NOON and entangled coherent states (ECS). We also discuss the scaling of the minimum phase uncertainty of EGCS for different Fock states that are displaced. In Sec.~\ref{SecIV}, we present an experimental scheme for the generation of EGCS (namely, with displaced single photon state) and finally summarize our work in Sec.~\ref{SecV}.
    \section{Minimum Phase Uncertainty for EGCS}\label{SecII}

    \begin{figure}[ht]
        \centering
        \includegraphics[width=0.5\textwidth]{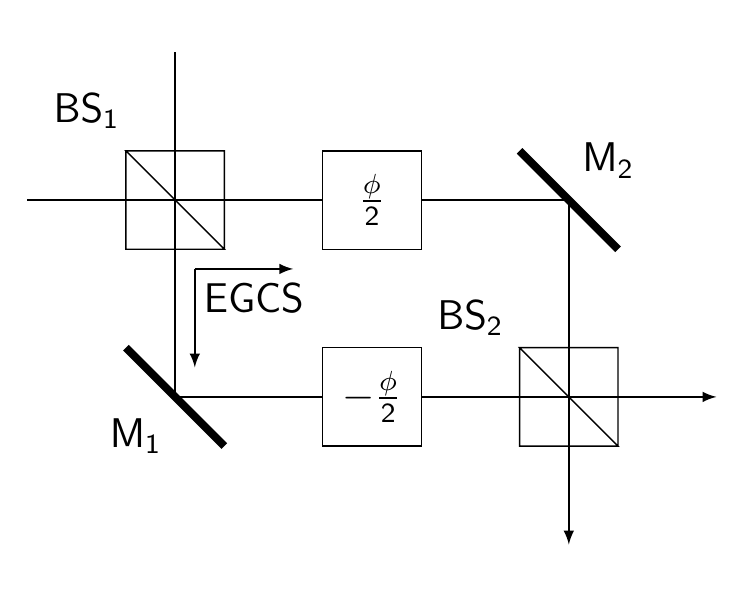}
        \caption{\textbf{Experimental scheme for photonic quantum metrology:} General interferometric setup for performing photonic quantum metrology measurements. BS$_i$, M$_i$ ($i=1$) and $\pm \phi/2$ represent 50:50 beam splitters, mirrors and phase differences in the two arms of the interferometer, respectively. EGCS is introduced in the interferometer (on choosing appropriate inputs to the interferometer) after the first beam-splitter BS$_1$ and evolves with the phase difference ($\pm \phi/2$). We describe the generation of this state in Sec.~\ref{SecIV} }
        \label{fig:schematic2}
    \end{figure}
    
    In photonic quantum metrology, single parameter estimation can be performed by measuring the change in phase in the two arms of an interferometer. The said parameter must be encoded in the phase difference between the interferometric paths. Now, let us assume that the interferometer with equal and opposite phases ($\pm\phi/2$) in the two arms [see Fig.~\ref{fig:schematic2}]. In experiments, this can be considered to be equivalent to having a phase difference ($\phi$) in one of the arms. The corresponding Hamiltonian depicting the phase-dependent evolution of a state through this section of the interferometer is given by: 
    \begin{equation}\label{Hamiltonian}
        H(\phi) = \phi H = \phi\frac{1}{2}(\hat{a}^\dagger\hat{a}-\hat{b}^\dagger\hat{b})
    \end{equation}
    where $\hat{a}$[$\hat{b}$] is the photon annihilation operator for output mode 1[2]. For a pure state, we can write the quantum Fisher information ($F^Q_\phi$) as~\cite{pang2014quantum}:
    
    \begin{equation}\label{Fisher_information}
        F^Q_\phi = 4t^2\bra{\Psi}\Delta H^2\ket{\Psi}
    \end{equation}
    where $t$ is the time for which the state is evolved under the influence of the Hamiltonian $H(\phi)$.
    Further, the quantum Cramer-Rao bound, which bounds the precision limit of an unbiased estimator by the relation:
    \begin{equation}\label{CramerRaoBound}
        \Delta \phi \geq \frac{1}{\sqrt{\mu F^Q_\phi}}
    \end{equation}
    For a single shot experiment, $\mu = 1$. Thus, using [Eqs.~\ref{Fisher_information} and~\ref{CramerRaoBound}] we can write:
    \begin{equation}
        \Delta \phi \geq \frac{1}{2\sqrt{\langle \Delta^2H \rangle}}
    \end{equation}
    Now, using the above equation, we proceed to calculate the minimum uncertainty in measuring $\Delta\phi$ for a particular input state ($\ket{\Psi}$) at the interferometer.

    We will analyze this uncertainty for an ECGS, which is formed after the first beam-splitter of the interferometer [shown in Fig.~\ref{fig:schematic2}]. We will use the following form of EGCS, which is entangled in the two output modes 1 and 2 of the first beam splitter of the interferometer (in Fig.~\ref{fig:schematic}), given by:
    
    \begin{equation}\label{EGCS}
        \ket{\Psi_{EGCS}} = N_{EGCS}(\ket{0}_1\ket{n,\alpha}_2+\ket{n,\alpha}_1\ket{0}_2)
    \end{equation}
    
    where $\ket{n,\alpha} = \hat{D}(\alpha)\ket{n}$ ($\hat{D}(\alpha)$ and $\ket{n}$ are the displacement operator and Fock state with photon number $n$~\cite{}) is the generalized coherent state and $N_{EGCS} = \frac{1}{\sqrt{2(1+(\alpha)^ne^{-|\alpha|^2/2}/n!)}}$ is the normalization factor. 
    
    Let us define the photon number operator in this case as:
    
    \begin{equation}\label{Navg}
        \hat{N} = \hat{a}^\dagger\hat{a} + \hat{b}^\dagger\hat{b} 
    \end{equation}
    
    For ECGS, we obtain the mean photon number to be:
    
    \begin{equation}
        \Bar{N} = \frac{n+|\alpha|^2}{\sqrt{2(1+\alpha^ne^{-|\alpha|^2/2}/n!)}}
    \end{equation}
    
    When $\ket{\Psi_{EGCS}}$ is passed through the unbalanced interferometer, 
    the phase information is encoded in the output state as:
    
    \begin{equation}\label{EGCS_phi}
        \frac{(\ket{0}_1\ket{n,\alpha e^{-i\phi/2}}_2+\ket{n,\alpha e^{i\phi/2}}_1\ket{0}_2)}{\sqrt{2(1+(\alpha)^ne^{-|\alpha|^2/2}/n!)}}
    \end{equation}
    
    Now, using [Eqs.~\ref{Hamiltonian} and~\ref{EGCS_phi}], we obtain $\langle\Delta^2H\rangle$ as:
    \begin{equation}\label{DelH}
        \langle\Delta^2H\rangle = \frac{n^2+|\alpha|^4+(4n+1)|\alpha|^2}{4(1+\alpha^ne^{-|\alpha|^2/2}/n!)}
    \end{equation}
    
    We see that in the limit $\Bar{N}>>|\alpha|^2$, we can approximate $\langle\Delta^2H\rangle \simeq \Bar{N}^2/2$. Thus, for such states, one can obtain better measurement precision than the shot-noise limit.
    
    \section{Comparison of $\Delta\phi$ of EGCS with other states}\label{SecIII}

    Previously, we considered the minimum phase uncertainty ($\Delta\phi$) that be achieved by entangled GCS when inputted at an interferometer. In this section, we compare the scaling of $\Delta\phi$ achieved by EGCS to that of other important states (NOON~\cite{cooper2010entanglement} and Entangled Coherent States~\cite{joo2011quantum}). To verify the scaling of the uncertainty in phase with the average photon number for all the mentioned states, we also perform numerical analysis which we describe in the following.
    
    \begin{figure}[ht]
        \centering
        \includegraphics[width=\linewidth]{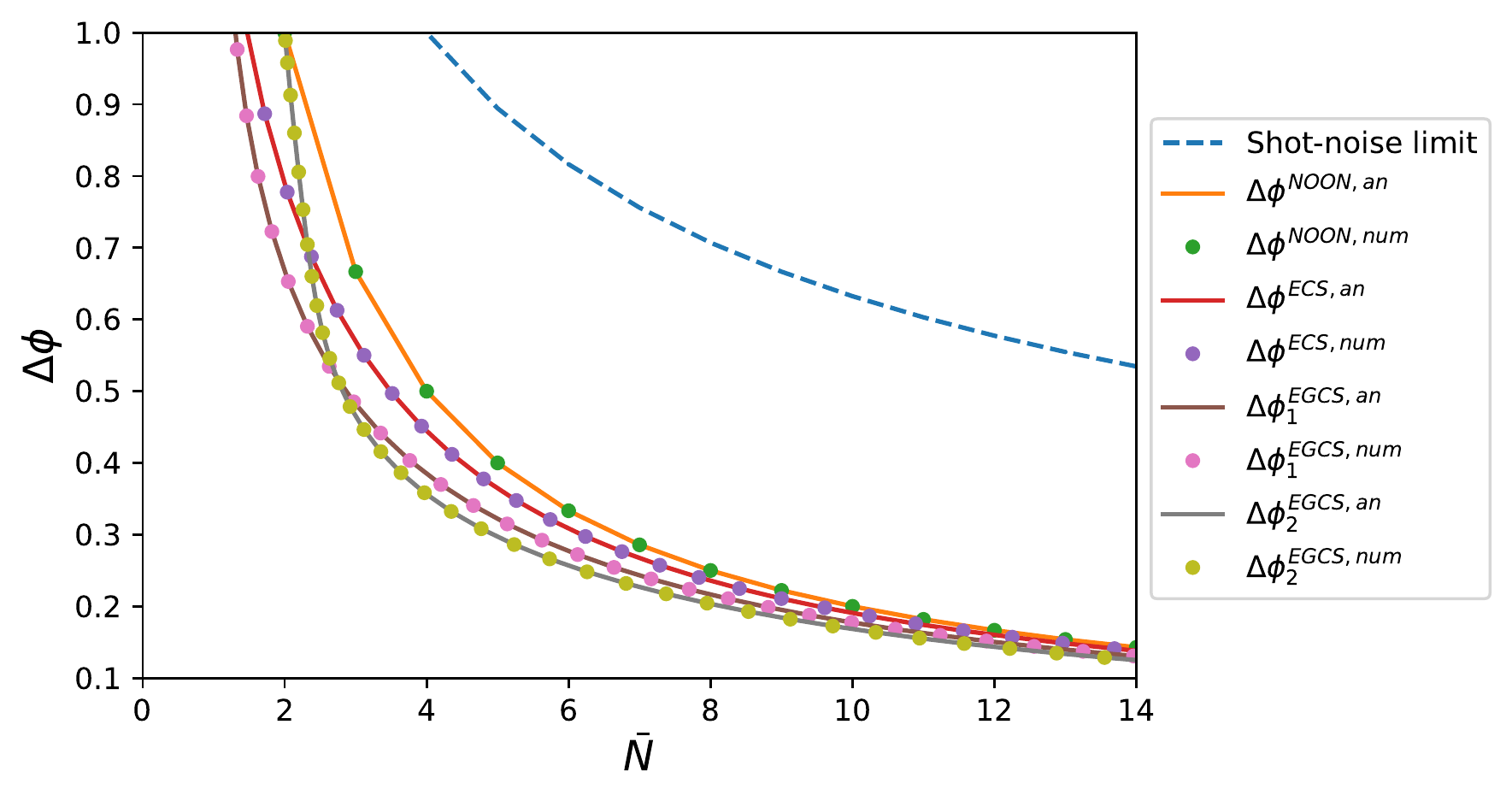}
        \caption{\textbf{Comparison of the phase estimation by various quantum states.} Plot of minimum phase uncertainty ($\Delta\phi$) vs. with average photon number ($\Bar{N}$) for NOON, entangled coherent and generalized entangled coherent states (up to $n=2$). $\Delta\phi^{NOON}, \Delta\phi^{ECS}, \Delta\phi^{EGCS}_1$ and $\Delta\phi^{EGCS}_2$ are the minimum phase uncertainties obtained by NOON state, ECS, EGCS for $n=1$ and $n=2$, respectively. The dots and lines represent numerical and analytical results (superscripted as "num" and "an" in the legend), respectively. The shot noise limit, i.e., $\Delta\phi\propto 1/\sqrt{\Bar{N}}$ is also plotted to show the advantage of phase sensitivity with the quantum states. It can be seen that the phase sensitivity for EGCS is better than the other states and increases with $n$.}
        \label{fig:comparison}
    \end{figure}
    
    For our numerical model, we use the Quantum Toolbox in Python (QuTiP) and the python packages $-$ \texttt{numpy} and \texttt{matplotlib}. We define a function to generate our state of interest in the general form: $\ket{n,\alpha}_1\ket{0}_2+\ket{0}_1\ket{n,\alpha}_2$ and then normalize it. When $n=0$ [$\alpha=0$], we obtain the entangled coherent state[NOON state], respectively. We then define a function to construct the number operator as given by [Eq.~\ref{Navg}] which can act on the generated state to produce the mean photon number $\Bar{N}$. As $\langle H\rangle = 0$ and $\langle\Delta^2 H\rangle = \langle H^2\rangle$, we define a function calculate $\langle H^2\rangle$. This generates the variance in $H$ from which we calculate the optimal phase estimation, $\Delta\phi_{min} = \frac{1}{2\sqrt{\langle\Delta^2H\rangle}}$. Now, we store the numerically obtained values for $\Delta\phi_{min}$ and $\Bar{N}$ for each of the considered states and plot them as shown in Fig.~\ref{fig:comparison}. We see that the numerically obtained values match with the analytically expected results. Note that while the plotted NOON state analytical function is represented by a continuous line, it is also calculated at discrete $\Bar{N}$ points. From Fig.~\ref{fig:comparison}, we observe that for larger $\Bar{N}$, $\Delta\phi^{EGCS}_n \simeq \Delta\phi^{ECS} \simeq \Delta\phi^{NOON}$ for $n = 1,2$, indicating that the ECS and EGCS are dominated by the NOON state amplitude at $\Bar{N}$. However for small $\Bar{N}$, 
    $\Delta\phi^{EGCS}_2 (\Bar{N}>2) < \Delta\phi^{EGCS}_1 (\Bar{N}>1) < \Delta\phi^{ECS} < \Delta\phi^{NOON}$, 
    thereby implying that the enhanced sensitivity reported using ECS can further be generalized. Hence, ECS is a part of a bigger class of states which we call EGCS where ECS is EGCS with $n=0$. Also, NOON state is another form of generalization of EGCS where $\alpha = 0$. For small $\Bar{N}$, such enhanced sensitivity in $\phi$ can be explained by the fact that $\Bar{N}^{EGCS}_n>\Bar{N}^{ECS}$. Further, EGCS contains the superposition of NOON states including those having $\Bar{N}>n+|\alpha|^2$, which contributes to higher sensitivity. A similar case can also be made for better sensitivity from ECS over the NOON state. Therefore, EGCS shows an advantage over ECS for low mean photon number, which is a preferable scenario in experiments as it is difficult to produce such states with high average photon numbers.
    
    We also observe [see Fig.~\ref{fig:comparison}] that EGCS with $n=1$ or $n=2$ has higher uncertainties in phase for $\Bar{N}<1$ and $\Bar{N}<2$ as compared to ECS and NOON state. As $\Bar{N}_{EGCS} = N_{EGCS}(n+|\alpha|^2)$, where $N_{EGCS}$ is the normalization factor defined earlier, for $\alpha = 0$, $\Bar{N_{EGCS}}|_{min} = N_{EGCS}n$. Thus, if $n\ne 0$, then $\Bar{N_{EGCS}}|_{min} \ne 0$. Therefore, we observe that for ECS and NOON state, the minima in $\Bar{N}$ can be set at 0, when $\alpha$ and $n$ are $0$ respectively. But for EGCS, $\Bar{N}$ is shifted by $n$ at $\alpha = 0$. Since we are considering a generalization of the better phase sensitivity obtained by ECS using EGCS, we consider varying $\alpha$ starting from $0$, keeping a fixed $n$. This shows the initial higher uncertainties in Fig.~\ref{fig:comparison} for EGCS with $n=1,2$. 
    
    \begin{figure}[ht]
        \centering
        \includegraphics[width=\linewidth]{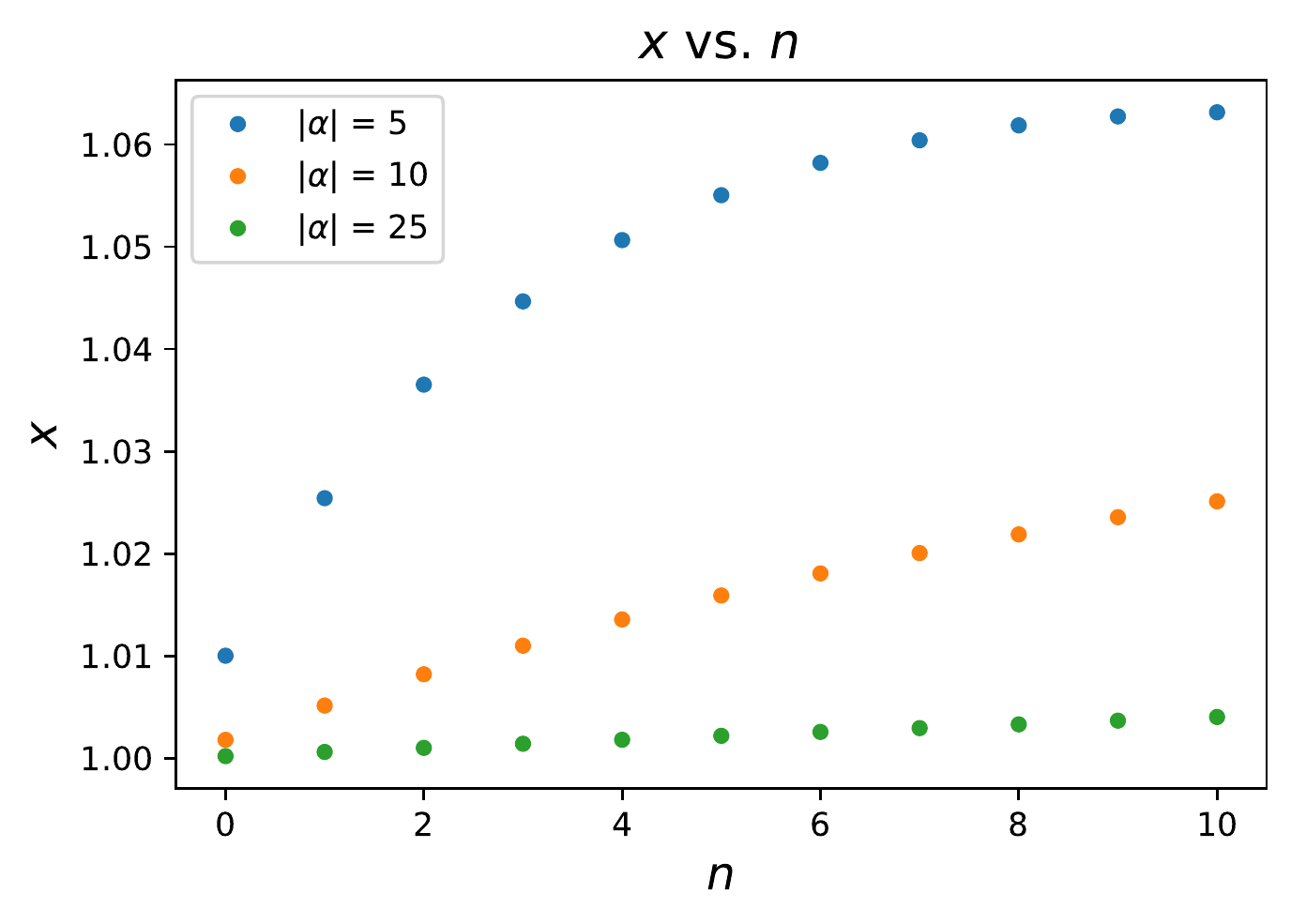}
        \caption{Scaling of the advantage with $n$ and $|\alpha|.$ Plot for $x$ vs. $n$ (Fock state photon number of the EGCS) where $x$ is the exponent from the fitting of $\Delta\phi$ with $1/\Bar{N}^x$ with varying $|\alpha|_{max}$. $\Bar{N}$ is varied by fixing $n$ and $\alpha$ varies from $(0,|\alpha|_{max})$. For $n=0$, ECS can be retrieved from the EGCS.}
        \label{fig:fitting}
    \end{figure}
    
    EGCS differ from ECS states by the Fock state that is displaced. Thus, to further elucidate the advantage of phase estimation using EGCS over ECS, we now check how the phase uncertainty ($\Delta\phi$) varies as a function of the average photon number ($\Bar{N}$). Thus, we fit our numerically generated $\Delta\phi$s, which follows the analytically derived function [shown in Fig.~\ref{fig:comparison}], with the corresponding $\Bar{N}$s. We vary the $\Bar{N}$ (see Eq.~\ref{Navg}) for the ECS and EGCS by fixing $n$ and varying $|\alpha|$ from $0$ to $|\alpha|_{max}$. Our fitting model obeys the equation: $\frac{1}{\Bar{N}^x}$, and $x=0.5$ or $1$ in the shot-noise and Heisenberg limits, respectively. 
    
    In Fig.~\ref{fig:fitting}, we plot the $x$ from our fitting equation with $n$, i.e., the photon number of $\ket{n}$ from Eq.~\ref{EGCS}. Further, our plot shows the variation of $x$ with $n$ for different $|\alpha|_{max}$. We see that for small $|\alpha|_{max}$ the EGCS exhibit a higher value of $x$ as compared to the ECS, and slightly higher than the Heisenberg limit. With $|\alpha|=5$, we observe that the value of $x$ saturates to $1.06$ for $n>8$, approximately. However for increasing $|\alpha|_{max}$, we see that all the states behave in a similar manner and $x$ increases very slowly with $n$, with a value close to $1$. 
    
    As we theorize in Sec.~\ref{SecII}, in the small $|\alpha|_{max}$ regime, the EGCS provide improved phase sensitivity (even beyond the Heisenberg limit) as compared to ECS. However, for large $|\alpha|_{max}$, all the states behave in a similar manner, as the amplitude of $\Bar{N}$ is dominated by $|\alpha|^2$ and hinders the effect of $n$. 
    
    \section{Entangled Generalized Coherent State Generation Scheme}\label{SecIV}
    
    \begin{figure}[ht]
        \centering
        \includegraphics[width=0.7\linewidth]{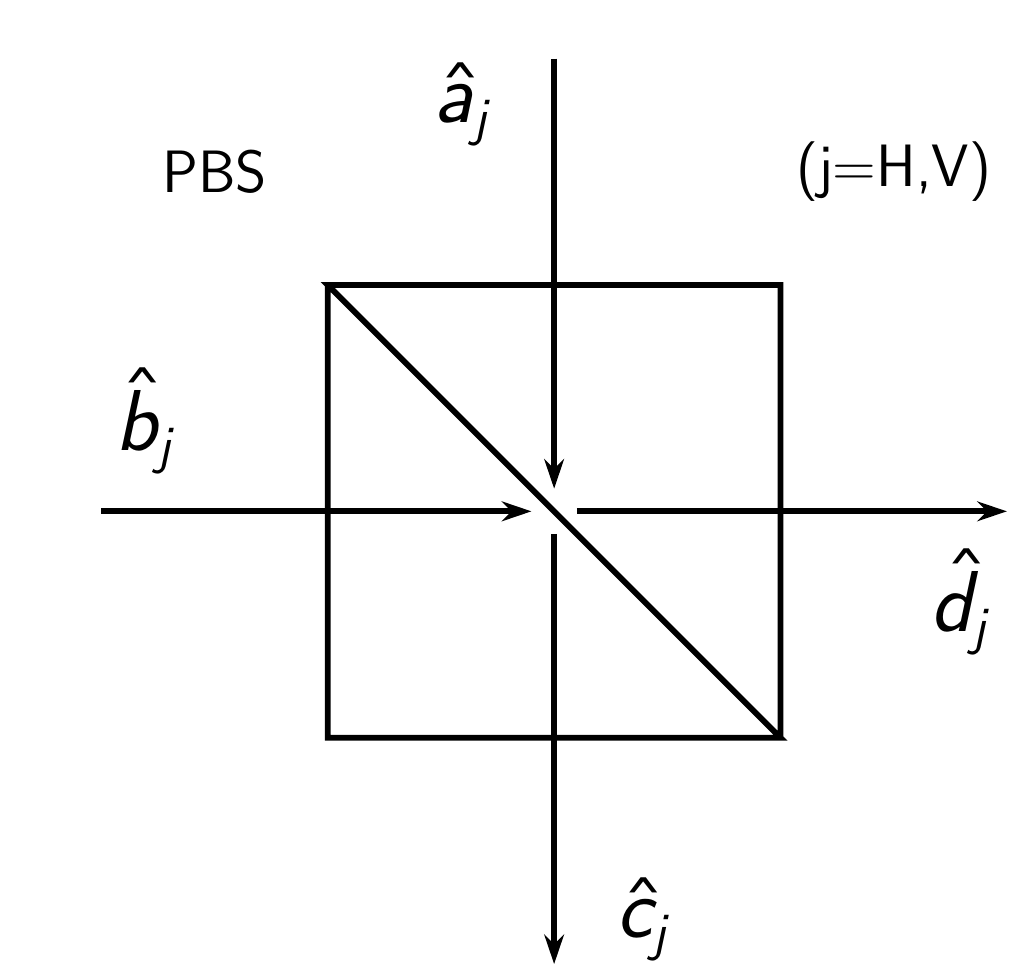}
        \caption{\textbf{Polarizing Beam Splitter.} Schematic diagram of a polarizing beam splitter (PBS) with input ports ($\hat{a}_j$, $\hat{b}_j$) and output ports ($\hat{c}_j$, $\hat{d}_j$), $j = H,V$ where $H, V$ signify horizontal and vertical polarization directions of light, respectively.}
        \label{fig:PBS}
    \end{figure}
    
    \begin{figure*}[ht]
        \centering
        \includegraphics[width=\textwidth]{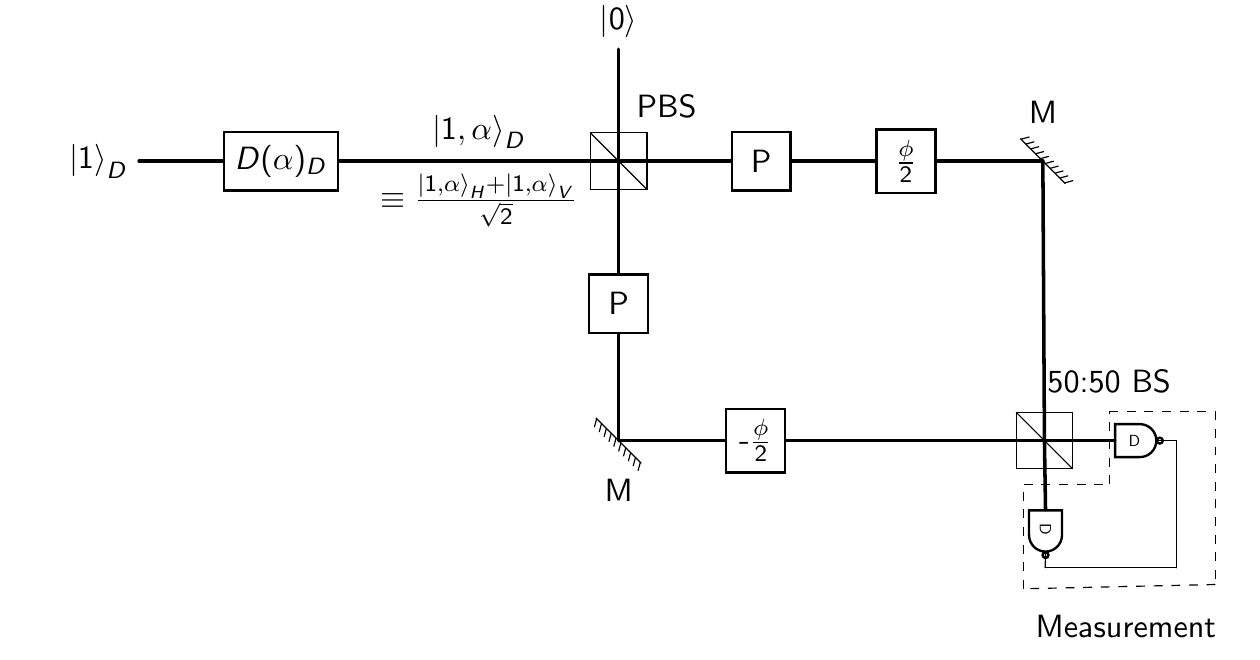}
        \caption{\textbf{Generation of EGCS.} The proposed experimental scheme to generate EGCS with $n=1$. $H,V$ and $D$ symbolize horizontal, vertical and 45$^o$ linear polarization directions of light. PBS, P, M and D signify polarizing beam splitter, polarizer oriented at 45$^o$ w.r.t the horizontal direction, mirror and detector, respectively. $\hat{D}(\alpha)_D$ can be applied to the input state $\ket{1}_D$ by using a Mach zehnder interferometer with one input fed by $\ket{\alpha}_D$ as described in~\cite{PhysRevA.105.062459}. }
        \label{fig:schematic}
    \end{figure*}
    Certain theoretical schemes have been suggested for using a Mach Zehnder interferometer (MZI), fed with a coherent state at one input, to act as a displacement operator~\cite{PhysRevA.105.062459}. This setting displaces an input state fed into the other input of the MZI, by a factor that depends on the product of the effective transmittivity of the MZI and amplitude of the coherent state. Now, using this we can produce a generalized coherent state of the form $\ket{n,T\beta} = \hat{D}(T\beta)\ket{n}$, where $T$ is the effective transmittivity of the MZI and $\beta$ is the amplitude of the input coherent state. 
    
    Among the Fock states, single photon states have been experimentally generated in various ways~\cite{eisaman2011invited} and thus, are more feasible to produce. Therefore, to produce $\ket{1,\alpha}$, where $|\alpha| = |T\beta|$, we can input a single photon state to a MZI with another input fed by a coherent state with amplitude $\beta$. Further as discussed in Sec.~\ref{SecIII}, we see an advantage of sensing protocols using entangled generalized coherent states such that with increasing $n$, the minimum uncertainty in measuring $\phi$ decreases. However, as single photon states are more feasibly produced using experimental setups, we address the generation of the EGCS: 
    
    \begin{equation}
        \ket{\psi^1_{EGCS}}=N_{1}(\ket{0}_1\ket{1,\alpha}_2+\ket{1,\alpha}_1\ket{0}_2)
    \end{equation}
    
    where $N_1 = \frac{1}{\sqrt{2(1+\alpha e^{-|\alpha|^2/2})}}$. 
    
    The states discussed above, are entangled in intensity. Thus, the source produces 2 beams (say beams A and B) among which $-$ 50\% of the times, only beam A[beam B] has non-zero intensity of the form $\ket{1,\alpha}$ and beam B[beam A] has the vacuum mode. The experimental generation of such states is quite tedious in general (see Appendix~\ref{A1}). We discuss an alternative here where entangling the modes with different polarization can be performed.
    
    First, we consider a single photon state in the $45^o$ linear polarization: $\ket{1}_D$. Now, the displacement operator can be applied to it by passing it through the MZI with a coherent state $- \ket{\alpha}_D$ (horizontal polarization). Thus the state undergoes the transformation as:
    
    \begin{equation}\label{EGCS_D}
        \ket{1}_D \xrightarrow[]{\hat{D}(\alpha)_D} \ket{1,\alpha}_D
    \end{equation}
    
    
    
    Next, we consider a polarizing beam splitter (PBS)~\cite{leonhardt2003quantum}. Let the input ports[output ports] be labelled as $a,b$[$c,d$] $-$ see Fig.~\ref{fig:PBS}. The photon annihilation operators for these ports are related as:
    
    \begin{equation}\label{PBS}
        \begin{pmatrix}
            \hat{c}_H\\\hat{c}_V\\\hat{d}_H\\\hat{d}_V 
         \end{pmatrix} \xrightarrow[]{PBS}
        \begin{pmatrix}
                0 & 1 & 0 & 0\\
                0 & 0 & 1 & 0\\
                1 & 0 & 0 & 0\\
                0 & 0 & 0 & 1
            \end{pmatrix}\begin{pmatrix}
            \hat{a}_H\\\hat{a}_V\\\hat{b}_H\\\hat{b}_V 
        \end{pmatrix}
    \end{equation}
    
    where the subscript $H$ and $V$ denote horizontal and vertical polarization, respectively. This denotes a polarizing beam splitter that transmits horizontally and reflects vertically polarized light.  
    
    We assume that the state given in Eqs.~\ref{EGCS_D} is incident on the input port, labelled $a$, of the polarizing beam splitter. Thus, the in the horizontal (H) and vertical (V) polarization basis, we can write the state ($\ket{\psi_{in}}$) as:
    \begin{equation}
        \ket{\psi_{in}} = \frac{\hat{D}(\alpha)_{H_a}\hat{a}_H^\dagger+\hat{D}(\alpha)_{V_a}\hat{a}_V^\dagger}{\sqrt{2}}\ket{0}_a\ket{0}_b 
    \end{equation}
     
     where, the other input port, labelled $b$, has the vacuum mode as an input. Also, $\hat{D}(\alpha)_i = e^{\alpha \hat{a}_i^\dagger-\alpha^*\hat{a}_i}$ where $i = H\text{ or }V$. 
    
    Considering the unitary operation applied by a polarizing beam splitter transforms the photon annihilation operators according to Eq.~\ref{PBS}, we get:
    
    \begin{equation}\label{PBS_fin_state}
        \ket{\psi_{in}} \xrightarrow[]{PBS} 
        \ket{1,\alpha}_{c_V}\ket{0}_d + \ket{0}_c\ket{1,\alpha}_{d_H}
    \end{equation}
    
    We drop the $\frac{1}{\sqrt{2}}$ factor as the state must be re-normalized after the PBS operation. Further, note that $\ket{0}_i = \ket{0}_{i_H}\ket{0}_{i_V}$[$i=c,d$] and for $\ket{1,\alpha}_{c_V}[\ket{1,\alpha}_{d_H}]$ there is also a $\ket{0}_{c_H}$[$\ket{0}_{d_V}$], which we have not written for brevity. The above state is a superposition of two product states in output modes $c$ and $d$, with different polarization. As these states need to be used in an interference experiment, the polarization must be the same for the entire combination of the modes in the composite state. Therefore, to resolve this, the state can be modified to have a $45^o$ linear polarization w.r.t the $x$ axis. This can be done by placing a polarizer (P) oriented at a $45^o$ angle with the $x$ axis, thereby converting the output state to have the same polarization.
    
    Thus, we can retrieve our desired EGCS (with $n=1$) state as:
    
    \begin{equation}\label{HWP_EGCS}
        \ket{1,\alpha}_{c_V}\ket{0}_d + \ket{0}_c\ket{1,\alpha}_{d_H} \xrightarrow[]{P \text{ at } 45^o} 
        \ket{1,\alpha}_{c}\ket{0}_d + \ket{0}_c\ket{1,\alpha}_{d}
    \end{equation}
    
    such that both terms of Eq.~\ref{HWP_EGCS} denote $45^o$ linearly polarized light. On comparing Eqs.~\ref{HWP_EGCS} and~\ref{EGCS}, we can say that Eq.~\ref{HWP_EGCS} is the EGCS for $n=1$ upto a normalization factor. Now this state is passed through two arms of an interferometer, with a $\phi$ phase difference between the two paths. We summarize the proposed experimental scheme in Fig.~\ref{fig:schematic}. This method can also be extended for EGCS with arbitrary $n$ by replacing $\ket{1}_D$ with $\ket{n}_D$ in Eq.~\ref{EGCS_D}.

    \section{Conclusion}\label{SecV}
    
    In summary, we have analytically and numerically shown that certain states such as entangled generalized coherent states show enhanced sensing beyond the classical limit. The entangled generalized coherent states are of the form $\ket{n,\alpha}\ket{0}+\ket{0}\ket{n,\alpha}$ (up to normalization), where $\ket{n,\alpha}$ is the Fock state $\ket{n}$ displaced by $\alpha$. These states, when passed through an interferometer with unbalanced arms (having a $\phi$ relative phase), can outperform the phase sensitivity limit of well-known ECS and NOON states under the same conditions and mean photon number. Further, we show that the sensitivity increases with increasing $n$, but saturates for a large value of $|\alpha|$. Therefore, these states show an advantage over the ECS in the small $\alpha$ regime. 
    
    Further, as  experimentally producing single photon states is more common than other Fock states, we also propose an experimental scheme to generate the entangled generalized coherent states for $n=1$. The scheme uses a Mach-Zehnder interferometer, a coherent state, polarizing beam splitter and a polarizer. This is a simplistic scheme for generating the EGCS which shows enhanced sensitivity over ECS $-$ a state that requires a more cumbersome experimental setup.
    
    The above results on phase sensitivity were obtained by considering a pure state to be inputted at the interferometer. In our future work, we plan to also explore the sensitivity under noisy experimental conditions, i.e., by considering a mixed state input.

    \appendix
    \section{EGCS Generation with Beam-Splitter}\label{A1}
    To produce the required EGCS (with $n=1$) using a beam splitter [having $a,b$ as input modes and $\hat{a}, \hat{b}$ as the photon annihilation operators, respectively], we assume the following input state:
    \begin{align}\label{input}
        \frac{1}{\sqrt{2}} \Bigg[\; {\ket{1,\frac{\alpha}{\sqrt{2}}}}_{a} \left( \; {\ket{\frac{\alpha}{\sqrt{2}}}}_{b} + {\ket{\frac{-\alpha}{\sqrt{2}}}}_{b} \right) + \\ 
        {\ket{\frac{\alpha}{\sqrt{2}}}}_{a} \; \left( \; {\ket{1,\frac{\alpha}{\sqrt{2}}}}_{b} + {\ket{1,\frac{-\alpha}{\sqrt{2}}}}_{b} \; \right) \Bigg]
    \end{align}
    This can be expanded into:
    \begin{align}
        \frac{1}{\sqrt{2}} \Bigg[ {\ket{1,\frac{\alpha}{\sqrt{2}}}}_a{\ket{\frac{-\alpha}{\sqrt{2}}}}_b - {\ket{\frac{\alpha}{\sqrt{2}}}}_a{\ket{1, \frac{-\alpha}{\sqrt{2}}}}_b + \\
        {\ket{1,\frac{\alpha}{\sqrt{2}}}}_a{\ket{\frac{\alpha}{\sqrt{2}}}}_b + {\ket{1,\frac{\alpha}{\sqrt{2}}}}_a{\ket{1,\frac{\alpha}{\sqrt{2}}}}_b \Bigg]
    \end{align}
    The above form can be written as a set of displacement and photon creation operators acting upon the vacuum states and thus can be written as:
    \begin{align}
        \frac{1}{\sqrt{2}}\Bigg[D_a\left(\frac{\alpha}{\sqrt{2}}\right)D_b\left(\frac{-\alpha}{\sqrt{2}}\right)\left( a^{\dag} - b^{\dag}\right) + \\ 
        D_a\left(\frac{\alpha}{\sqrt{2}}\right)D_b\left(\frac{\alpha}{\sqrt{2}}\right)\left( a^{\dag} + b^{\dag}\right)\Bigg]\ket{0}_a\ket{0}_b 
    \end{align}
    Applying the transformation of the annihilation and creation operators as a result of passing through the beam splitter results in the following:
    \begin{align}
        \left[D_c(\alpha)\;c^\dag + D_d(\alpha)\;d^\dag\right]\ket{0}_c\ket{0}_d \Rightarrow \ket{1, \alpha}_a\ket{0}_b + \ket{0}_a\ket{1, \alpha}_b
    \end{align}
    
    which is the same as the EGCS ($n=1$) up to a normalization factor. However, as shown in Eq.~\ref{input}, the input state requires superpositions of two-mode states such that one mode contains displaced coherent state and the other has Schrodinger cat state with a coherent state in one mode and with displaced cat state in the other. Producing such two-mode states has not be explored yet, and thus, this is a more cumbersome scheme. Thus, in Sec.~\ref{SecIV}, we have provided an alternate way to produce these states and use them for interferometry for enhanced sensing.
    
    

%

\end{document}